\documentclass{mem}
\usepackage{natbib}
\usepackage{txfonts}
\usepackage{balance}
\usepackage{graphicx}
\usepackage{flushend}
\usepackage{color}
\usepackage[breaklinks,pdftex]{hyperref}

\idline{75}{282}
\begin{document}
\def\teff{$T\rm_{eff }$}
\def\kms{$\mathrm {km s}^{-1}$}

\title{
Revisiting the SFR-Mass relation at z=0 with detailed deep learning based morphologies
}

   \subtitle{}

\author{
H. \, Domínguez Sánchez\inst{1},
M. \, Bernardi\inst{2} and
M. \, Huertas-Company\inst{3, 4, 5}
}

\institute{Centro de Estudios de Física del Cosmos de Aragón, Plaza San Juan 1, 44001, Teruel, Spain
\and
Department of Physics and Astronomy, University of Pennsylvania, Philadelphia, PA 19104, USA
\and
Instituto de Astrof\'isica de Canarias, E-38200, La Laguna, Tenerife, Spain  
\and
LERMA, Observatoire de Paris, PSL Research University, CNRS, Sorbonne Universit\'es, UPMC Univ. Paris 06,
F-75014 Paris, France
\and
University of Paris Denis Diderot, University of Paris Sorbonne Cit\'e (PSC), 75205 Paris Cedex 13, France\\
\email{hdominguez@cefca.es}
}

\authorrunning{Domínguez Sánchez}

\titlerunning{Revisiting the SFR-M$^{*}$-morphology relation}

\date{Received: Day Month Year; Accepted: Day Month Year}

\abstract{
Galaxy morphology is a key parameter in galaxy evolution studies. The enormous number of galaxies which future and current surveys will observe demand of automated methods for morphological classification. Supervised learning techniques have been successfully used for the morphological classification of galaxies from different datasets, including Sloan Digital Sky Survey (SDSS), Mapping Galaxies with Apache Point Observatory (MaNGA) or Dark Energy Survey (DES). With these proceedings, we release the morphological catalogue for a sample of 670,000 SDSS galaxies based on the deep learning models trained on SDSS RGB images with morphological labels from human-based classification catalogues. The released catalogue includes binary classifications  (early-type versus late-type, elliptical versus lenticular, identification of edge-on and  barred galaxies) plus  a T-Type. The classifications also include k-fold based uncertainties. This is, as of today, the largest  catalogue including a T-Type classification. As an example of the scientific potential of this classification, we show how the location of the galaxies in the star formation - stellar mass  plane (SFR-M$^{*}$) depends on morphology. This is the first time the SFR-M$^{*}$ relation is combined with T-Type information for such a large sample of galaxies.

}

\maketitle{}

\section{Introduction}

Galaxy morphology is one of the key parameters in galaxy evolution studies. While the existence of an ordered sequence of galaxy appearance is well know since the beginning of the last century \citep{Hubble1926} its origin is still highly debated.
Galaxy morphology is strongly correlated with their stellar populations, but its connection with  mass assembly mechanisms  and quenching events is still unclear \citep[e.g.,][]{Hirschmann2015,Nelson2016,RodriguezGomez2016}. In order to shed some light on the interrelation between galaxy morphology and evolutionary paths, large samples of galaxies with robust morphological classifications at different cosmic epochs are needed. With the arrival of large imaging surveys, visual classification of galaxies becomes  unfeasible and automated methods are required.

Supervised deep learning (DL) methods based on Convolutional Neural Networks (CNN) using galaxy images as input has demonstrated to be  very successful for the classification of nearby bright galaxies for which large samples of previously labelled galaxies, such as Galaxy Zoo \citep{Willett2013} or \cite{Nair2010}, were available. In \cite{DS2018} we trained a CNN, paying special attention to the training sample selection (i.e., using only galaxies with large agreement among Galaxy Zoo classifiers) and we published what was, at the time, the largest DL-based morphological  catalogue, including 670,000 SDSS DR7 \citep{Abazajian2009} galaxies from the \cite{Meert2015} sample ($14 < \rm m_r < 17.7$).

In \cite{DS2022} we presented an improved version of the classification obtained in \cite{DS2018}. We used a vanilla convolutional neural network (CNN), consisting of four convolutional layers with
squared filters of different sizes (6, 5, 2, 3) followed by  dropout and 2$\times$2 \textit{maxpooling}. A fully connected layer returns one output value. The input are RGB  SDSS-DR7 cutouts  with a variable size  proportional to the Petrosian radius of the galaxy (5$\times$R$_{90}$). The cutouts are  re-sampled to 69$\times$69 pixels before being fed to the CNN. The RGB images were normalized to the maximum of each band to avoid any dependence of the morphological classification on color information. We used the \cite{Nair2010} catalogue to train a regression model which returns a T-Type (analogue to the Hubble sequence), and two binary models: one that separates early (ETG) or late type galaxies (LTG) and the other that separates elliptical (Es) from lenticular galaxies (S0)\footnote{This model is only meaningful for galaxies with T-Type $<$ 0.}. The low end of the T-Types was better recovered than in the previous version (Figure 4 in \citealt{DS2022}). The separation between ETGs and LTGs complements the T-Type classification, especially at the intermediate types (-1 $<$ T-Type $<$ 2), where the T-Type values are more uncertain.  The Galaxy Zoo catalogue  \citep{Willett2013} was used for training two binary models, one that identifies barred galaxies and another that identifies edge-on galaxies. In addition, k-fold-based uncertainties on the classifications were also provided. These models were applied to the MaNGA \citep{Bundy2015} DR17 final sample \citep{sdssDR17}, including $\sim$ 10,000 galaxies, and released in the form of the MaNGA Deep Learning Morphological DR17 Value Added Catalogue (MDLM-VAC-DR17)\footnote{\url{https://www.sdss4.org/dr17/data_access/value-added-catalogs/?vac_id=manga-morphology-deep-learning-dr17-catalog}}.

\begin{figure*}

\includegraphics[width=0.49\linewidth]{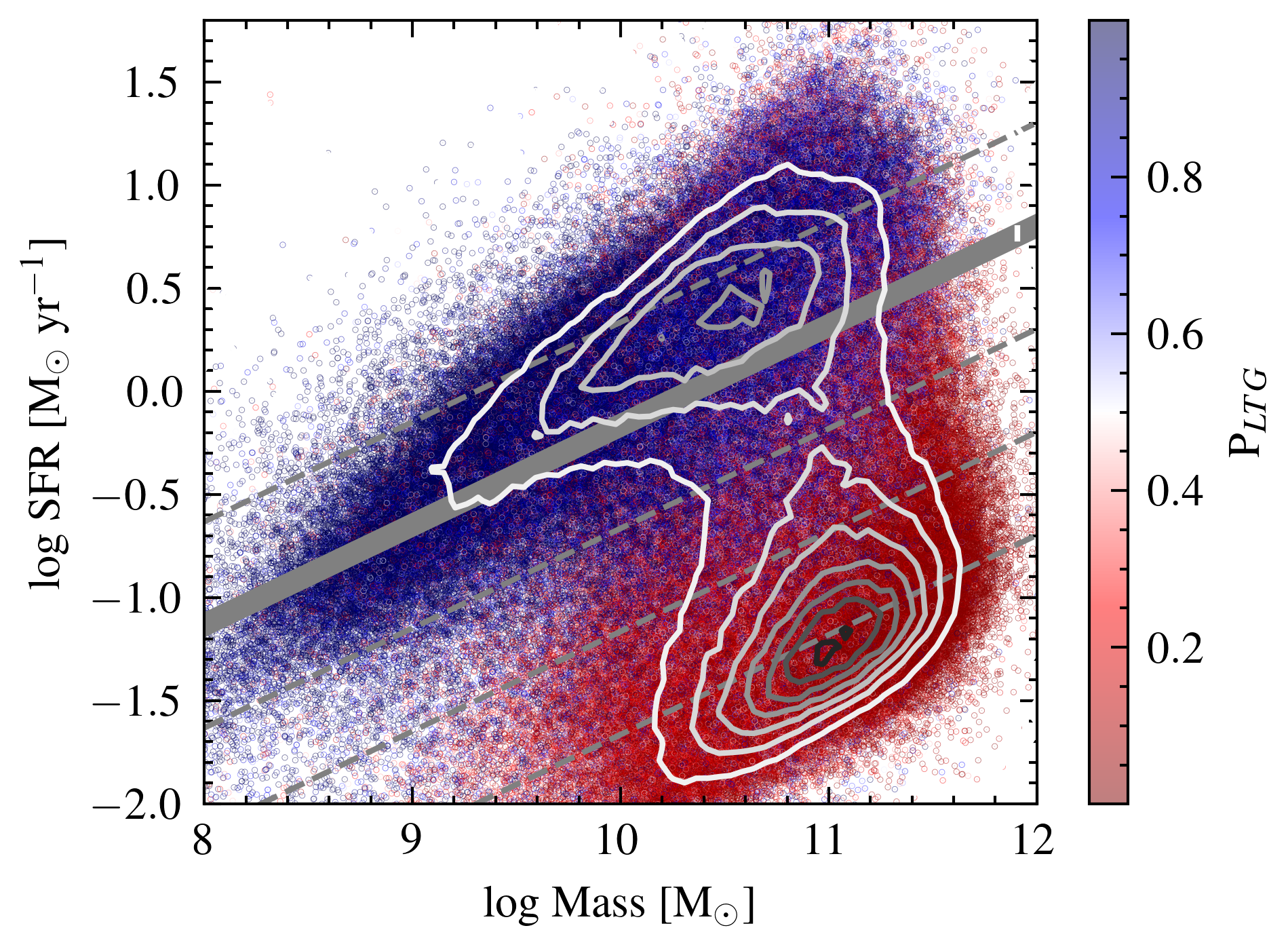}
\includegraphics[width=0.49\linewidth]{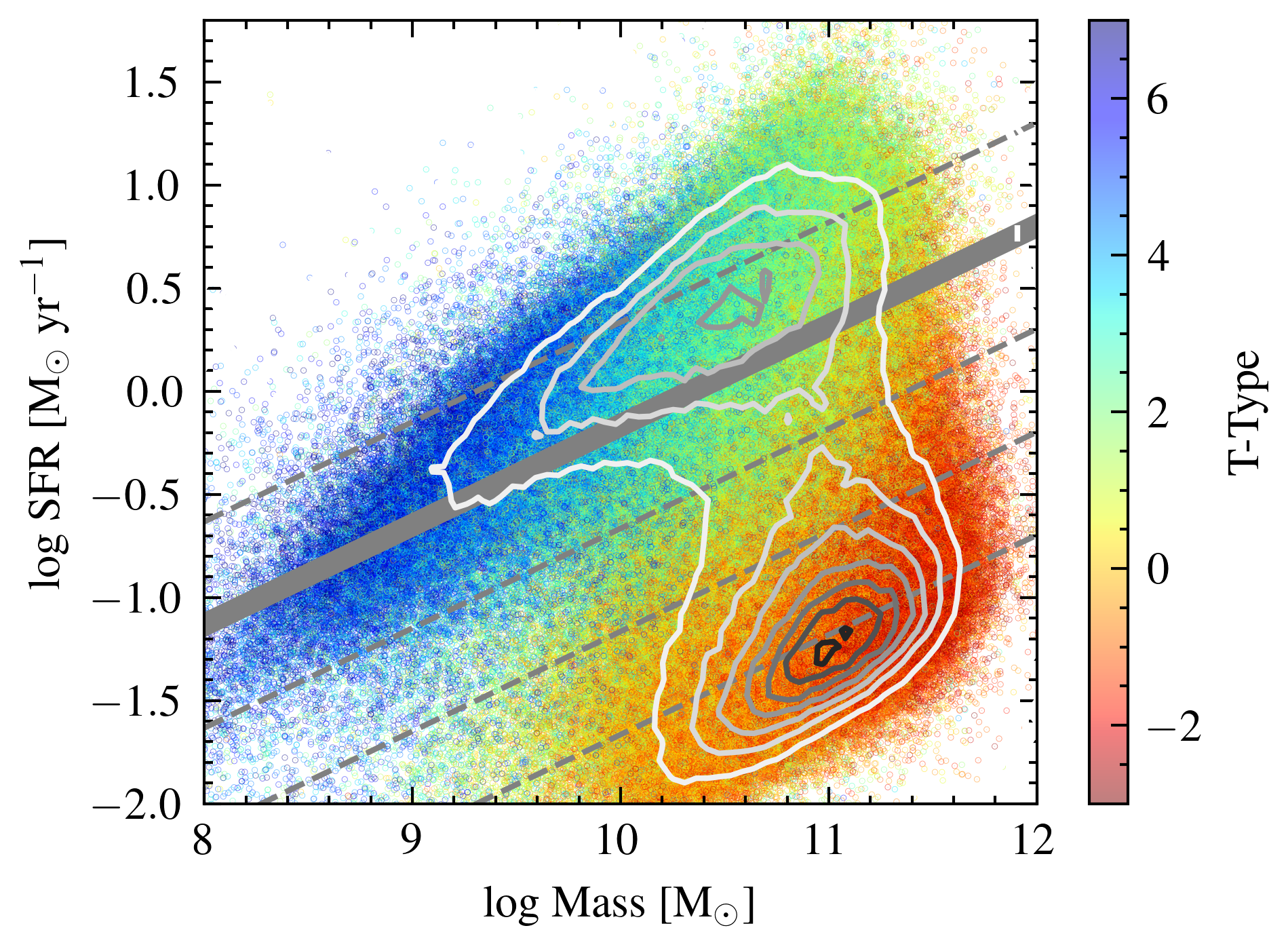}
  
\caption{
\footnotesize
Star formation rate versus stellar mass for a sample of 653,543 galaxies, color coded according to their probability of being LTG (left panel) and their T-Type (right). The thick grey line is the Main Sequence from \cite{Speagle2014} for the local Universe and the dashed grey lines are their shifted  versions by 0.5 dex.  Contours are over-plotted to highlight the most populated regions. No smoothing is applied to the colors. The T-Type unveils a much more complex view of the SFR-M$^*$ plane than a commonly used ETG/LTG separation.
}
\label{fig:SFR-Mass}
\end{figure*}

\section{Updated SDSS Morphological catalogue}
\label{sect:cat}

MaNGA is an Integral Field spectroscopic survey which provides resolved spectral information for each galaxy. However, the morphological classification is obtained by training the DL models with RGB SDSS images, meaning that MaNGA played no role at all in the construction of the morphological catalogue, except for the sample selection. We  have now applied the DL models from \cite{DS2022} to the full \cite{Meert2015} sample and we take the opportunity to release this new catalogue with these proceedings.  A detailed description of the construction of the models and the performance of the different classification tasks can be found in \cite{DS2022}. Since the imaging data and the magnitude range of the MaNGA DR17 and the \cite{Meert2015} samples are similar, we do not expect significant differences in the results.

The catalogue provides binary classifications (ETG vs LTG, E vs S0, edge-on, bars) and  can be found in this link \footnote{\url{https://archive.cefca.es/ancillary_data/sdss_morphological_catalogues/sdss_morphological_catalogues.tar.gz}}. Its content is identical to the MDLM-VAC-DR17, except for the visual classification (VC) and visual flag (VF), unfeasible for such a large galaxy sample as the one presented here. We refer the reader to Table 4 of \cite{DS2022} for a detailed description of the catalogue columns.

We recommend the following criteria for selecting samples of Es, S0 and spirals (S):

\begin{itemize}
    \item E: (P$_{\rm LTG}$ $<$ 0.5) and (T-Type $<$ 0) and (P$_{\rm S0}$ $<$ 0.5) 
    \item S0: (P$_{\rm LTG}$ $<$ 0.5) and (T-Type $<$ 0) and (P$_{\rm S0}$ $\ge$ 0.5) 
    \item S: (P$_{\rm LTG}$ $\ge$ 0.5) and (T-Type $\ge$ 0) 
\end{itemize}

where P$_{\rm LTG}$ separates ETGs from LTGs and P$_{\rm S0}$ separates Es from S0 (only meaningful for ETGs). 
Note that this is the most restrictive criteria, as it combines the information of the LTG/ETG classification with the T-Type. The thresholds at P$_{\rm LTG}$=0.5 and P$_{\rm S0}$=0.5  are a good compromise between completeness and purity (see Figure 5 and 7 in \citealt{DS2022}) but can be modified in order to obtain a more pure or complete S0 sample, depending on the users purpose. The above selection returns 18, 20 and 50\% of Es, S0 and S, respectively, leaving 12\% of the galaxies with an ambiguous classifications (their P$_{\rm LTG}$ and T-Type values are discordant). Alternatively, one can use the T-Type information only (which returns 18, 20 and  62~\% of E, S0 and S) or the P$_{\rm LTG}$ (which returns 18, 32, 50\%). The S0 is the population more affected by the different selection criteria, as already discussed in Section 3.4.1 of \cite{DS2022}.

\section{Scientific application: the SFR-Mass plane}
\label{sect:SFR-M}

\begin{figure*}[t!]
\resizebox{\hsize}{!}{\includegraphics[clip=true]{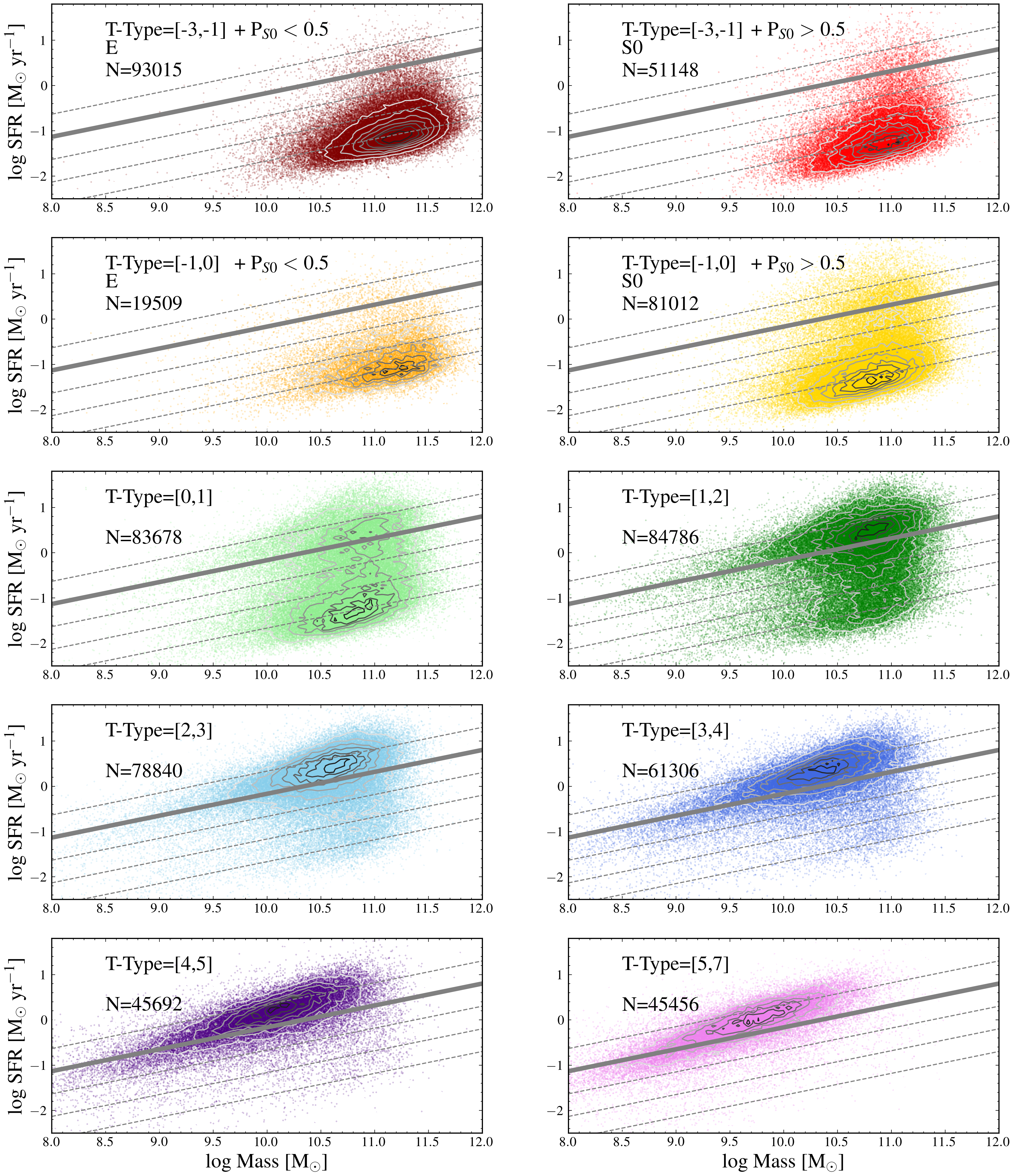}}
\caption{\footnotesize
Star formation rate versus stellar mass for galaxies divided in narrow T-Type bins, as reported in the legend. The two upper  panels separate elliptical galaxies  (selected as P$_{\rm S0}$ $<$ 0.5, left) from lenticulars (P$_{\rm S0}$ $\ge$ 0.5, left). The number of galaxies in each panel is reported. Grey lines are the same as in Figure \ref{fig:SFR-Mass} and the contours correspond to the distribution of the galaxies shown in each panel. 
}
\label{fig:bins}
\end{figure*}

It is well known that galaxy morphology is related to galaxy properties, in particular to stellar mass and star formation efficiency.  As an example of the scientific return of the morphological classification provided in the catalogue, we analyze the relation between morphology and the SFR-M$^*$ plane in this Section.

In Figure \ref{fig:SFR-Mass} we show the SFR-M$^*$ plane color coded by two of the classifications reported in the catalogue: P$_{\rm LTG}$ and T-Type. The SFR and M$^*$ values are retrieved from the MPA-JHU Stellar mass catalogue\footnote{\url{https://www.sdss4.org/dr17/spectro/galaxy_mpajhu}} (The Max Planck for Astrophysics and Johns Hopkins University groups), which provides galaxy properties for all DR8 galaxy spectra. Stellar masses are based on the $ugriz$ galaxy photometry and are calculated using the Bayesian methodology and model grids described in \cite{Kauffmann2003}. SFRs are computed within the galaxy fiber aperture (3\arcsec) using the nebular emission lines as described in \cite{Brinchmann2004}. SFRs outside the fiber are estimated using the galaxy photometry following \cite{Salim2007}. For AGN and galaxies with weak emission lines, SFRs are estimated from the photometry. There are 653,543 galaxies with reliable M$^*$ and SFR estimates (97\% of the galaxies included in our morphological catalogue).

In the left panel of Figure \ref{fig:SFR-Mass}, galaxies are color coded according to P$_{\rm LTG}$, i.e., the probability of a galaxy to be LTG rather than ETG. As expected, LTG galaxies are located in and above the main sequence (MS), while quenched galaxies show morphologies consistent with ETGs. A basic separation between elliptical/S0 and spirals is the most commonly classification reported in morphological catalogues.\footnote{Galaxy Zoo separates galaxies into \textit{`smooth'} or \textit{`features or disc'}, which is usually used as a proxy for the separation between ETGs and LTGs. Note, however, that being \textit{`smooth'} is not equivalent to being ETG and the contamination of \textit{`smooth'} galaxies by LTGs can be significant - see Figure 15 from \cite{DS2022}} While P$_{\rm LTG}$ provides a broad separation between two classes, the T-Type parameter, corresponding to the  Hubble sequence (or de Vaucouleurs type, \citealt{deVaucouleurs1963}) shows a more detailed and complex representation of the SFR-M$^*$ plane (right panel of Figure \ref{fig:SFR-Mass}). 

Galaxies with the lowest T-Types (T-Type $<$ 0, reddish colors) populate the high-mass and low SFR region (analogue to the red sequence in the color magnitude diagram) and the opposite happens for the galaxies with the largest T-Type values (T-Type $>$ 4, dark blue colors). Galaxies with intermediate T-Types populate the green valley but also the high-mass starburst region (above the MS) and the low-mass end of the quenched population. This is, to the best of our knowledge, the first time the SFR-M$^*$ is combined with T-Type information for such a large sample of galaxies. Note that  no smoothing is applied to the figure, hence the underlying relation between mass, star formation efficiency and structure naturally emerges.

To shed more light on how the T-Type correlates with the SFR-M$^*$ loci, Figure \ref{fig:bins} dissects the diagram in narrow T-Type bins. In addition, the upper panels separate elliptical (E) and lenticular (S0) galaxies according to their P$_{\rm S0}$ - we remind the reader that, although  P$_{\rm S0}$ is reported for all the galaxies in the catalogue,  it is only meaningful for galaxies with T-Type $<$ 0. 

Several clear trends turn up from this novel representation. The four upper panels show the distribution of galaxies with T-Type $<$ 0 (corresponding to ETGs), divided into E (left) and S0 (right). These galaxies are the more massive and have the lowest SFRs, as expected. The contours are concentrated in a relatively narrow region ($\sim$ 1 dex), which could be an analogue of the star forming MS for the quenched galaxies (QMS). It is worth noticing that the Es with T-Type=[-1,0] are less abundant than Es with T-Type=[-3,-1] and occupy a very narrow region in the SFR-M$^*$ plane, while  the S0s expand over a  wider SFR range.


Galaxies with intermediate T-Types (T-Type=[0, 2], green colours) expand through a large SFR range ($\sim$~3~dex) and show a bimodal distribution, with galaxies with T-Type=[0, 1] being more abundant in the low SFR region than galaxies with T-Type=[1, 2]. This could be interpreted as the existence of two distinct galaxy populations with Sa/Sab morphologies, or, alternatively, could  be an indication that these  galaxies are being quenched and we are witnessing their evolutionary tracks as they cross the green valley. More detailed studies regarding their ages and star formation histories should be carried out to support this statement beyond speculation.

Finally,  galaxies with  T-Types $>$ 2 (corresponding to Sb, Sc, Sd), are mostly located above the MS, with less and less galaxies below the MS as we move to lager T-Type values. There is also an evident shift towards lower masses and a narrowing of the location of the galaxies with increasing T-Type, with a slightly steeper slope than the MS. We remark that size, mass and colour played no role in the morphological classification, which was purely based on SDSS imaging.

\section{Towards the classification of high redshift galaxies}

The success of DL for classifying large samples of galaxies is undeniable. However, one of the main drawbacks of supervised deep learning is that they need large samples of labelled galaxies. In addition, they are strongly affected by domain shifts, whether caused by instrumental effects or by  different parameter space distribution of the galaxy properties. This a big challenge for classifying high redshift galaxies, which are usually much fainter than their lower redshift counterparts. 

One way to overcome the lack of a large training sample is the use of `transfer learning', i.e., using the weights learned by a model for a particular data set for initializing the training with new data, rather than using a random initialization. In \cite{DS2019} we adapted the SDSS models to the DES data, demonstrating that this approach allows to reduce the size of the training sample by one order of magnitude. In \cite{VegaFerrero2021}, we were able to classify galaxies much fainter (m$_r$  $<$ 22) than the ones with available labels (m$_r$  $<$ 17.7) by  `emulating' how the local galaxies would look like at higher redshifts, while keeping their original labels for training. The classifications where highly accurate (accuracy=97\%) and their performance was consistent throughout all the magnitude range. The corresponding catalogue, including 27 million galaxies, was released with the paper and can be found here\footnote{\url{https://des.ncsa.illinois.edu/releases/y3a2/gal-morphology}}. Unfortunatley, the image resolution was not enough for providing a T-Type classification and only allowed for a basic ETG/LTG separation and the identification of edge-on galaxies. 

Alternative methods which do not require of labelled samples, such as self-supervised learning (e.g., \citealt{Sarmiento2021}) or Principal Component Analysis (e.g \citealt{Tous2022}) also provide valuable insights on galaxy properties. Finally, there are some tasks which still remain challenging for CNNs, for instance, the detection of low surface brightness structures like tidal features (see Domínguez Sánchez et al. 2023).

\section{Conclusions}

With these proceedings we release the morphological catalogue for the \cite{Meert2015} sample, based on the models presented in \cite{DS2022}. The catalogue provides binary classifications (ETG vs LTG, E vs S0, edge-on, bars) and a T-Type for $\sim$670,000 galaxies, being the largest sample up to date with such detailed morphological properties. The scientific potential of the catalogue is illustrated by dissecting the SFR-M$^*$ plane in narrow T-Type bins. The results highlight the strong dependence of SFR and mass with galaxy structure and suggest that the SFR main sequence depends on morphology. We leave for forthcoming studies a more robust statistical analysis of this evidence. Other important relations, such as the Size-Mass relation, or the fundamental plane should be reviewed, dissecting galaxies according to their T-Types. Finally, the role of bars in secular evolution will surely benefit from such a large sample of barred galaxies, while the identification of edge-on galaxies can be useful for several scientific purposes, from  estimating dust attenuation \citep{Masters2010} to probing of self-interacting dark matter \citep{Secco2018}.

\begin{acknowledgements}
HDS acknowledges the financial support by the PID2020-115098RJ-I00 grant from MCIN/AEI/10.13039/501100011033 and from the Spanish Ministry of Science and Innovation and the European Union - NextGenerationEU through the Recovery and Resilience Facility project ICTS-MRR-2021-03-CEFCA and by the PID2020-115098RJ-I00 grant from MCIN/AEI/10.13039/501100011033. MHC acknowledges financial support from the State Research Agency (AEI-MCINN) of the Spanish Ministry of Science and Innovation under the grant and “Galaxy Evolution with Artificial Intelligence” with reference PGC2018-100852-A-I00, from the ACIISI, Consejería de Economía, Conocimiento y Empleo del Gobierno de Canarias and the European Regional Development Fund (ERDF) under grant with reference PROID2020010057, and from IAC project P.301802, financed by the Ministry of Science and Innovation, through the State Budget and by the Canary Islands Department of Economy, Knowledge and Employment, through the Regional Budget of the Autonomous Community. The authors gratefully acknowledge the computer resources at Artemisa, funded by the European Union ERDF and Comunitat Valenciana as well as the technical support provided by the Instituto de Física Corpuscular, IFIC (CSIC-UV).

\end{acknowledgements}

\bibliographystyle{aa}
\bibliography{bonifacio}

\end{document}